\documentclass[twocolumn,prb,showpacs,preprintnumbers,amsmath,amssymb]{revtex4}

\usepackage{graphicx}
\usepackage{dcolumn}
\usepackage{bm}

\begin{document}


\title{In-plane and transverse superconducting fluctuation diamagnetism in the presence of charge-density waves in 2H-NbSe$_2$}

\author{F. Soto$^{1}$}
\author{H. Berger$^{2}$}
\author{L. Cabo$^{1}$}
\author{C. Carballeira$^{1}$}
\author{J. Mosqueira$^{1}$}
\author{D. Pavuna$^{2}$}
\author{F. Vidal$^{1}$}

\affiliation{$^{1}$LBTS, Departamento de F\'isica da Materia Condensada, Universidade de Santiago de Compostela, E-15782 Santiago de Compostela, Spain \\
$^{2}$Department of Physics, Ecole Politechnique F\'ed\'erale de Lausanne, CH-01015, Lausanne, Switzerland}

\date{\today}

\begin{abstract}

The fluctuation-diamagnetism (FD) above the superconducting transition was measured in 2H-NbSe$_2$ single crystals. The moderate uniaxial anisotropy of this compound, and some experimental improvements, allowed to measure the superconducting fluctuation effects in the two main crystallographic directions. These results reveal that the nonlocal electrodynamic effects on the FD are highly anisotropic, and they also discard a possible contribution to the FD coming from the \textit{charge-density waves} (CDW) appearing below $T_{\rm CDW}>T_C$ in 2H-NbSe$_2$, in agreement with a phenomenological estimate. 

\end{abstract}

\maketitle

\section{Introduction}

Above but near a superconducting transition, the normal state magnetization decreases due to the presence of evanescent precursor Cooper pairs created by the unavoidable thermal agitation.\cite{Tinkham} This effect, called fluctuation-induced diamagnetism (FD), was predicted by Schmidt\cite{Schmidt} and Schmid\cite{Schmid}, and first observed by Tinkham and coworkers in low-$T_C$ superconductors.\cite{Gollub} In contrast with most of the other superconducting fluctuation effects, the FD is not only proportional to the density of precursor Cooper pairs but also to some extent to their size, \textit{i.e.}, to the superconducting coherence length amplitude $\xi(0)$.\cite{Tinkham,Schmidt,Schmid,Gollub} This leads to a quite large FD amplitude relative to the normal-state magnetization in both low- and high-$T_C$ superconductors. As a consequence, in addition to its intrinsic importance, the FD has become a powerful tool to probe the superconducting transition in different materials, and in the last years it has been extensively used in high temperature cuprate superconductors (HTSC).\cite{Reviews,Cabo06}

In spite of its wide interest, an important aspect of the fluctuation-induced diamagnetism remains still open: the FD anisotropy and its interplay with the nonlocal electrodynamic effects. In particular, the \textit{transverse} FD (i.e., with the magnetic field $H$ applied parallel to the \textit{ab} crystallographic planes) has not been yet measured in any (quasi)uniaxial superconductor like, for instance, the layered ones. This is because for this direction, the FD is reduced by a factor $\gamma^2\equiv m_c^*/m_{ab}^*$ with respect to the \textit{in-plane} FD (i.e., with $H$ perpendicular to the \textit{ab} planes), being $m_c^*$ and $m_{ab}^*$ the transverse and in-plane effective masses. Just as an example, for optimally doped YBa$_2$Cu$_3$O$_{7-\delta}$ (one of the less anisotropic HTSC), $\gamma^2\approx$ 50-100.\cite{gammaYBCO} So, the measurement of transverse fluctuation effects constitutes a serious experimental challenge, because the small signal may easily be overcome by extrinsic $T_C$ inhomogeneities, uncertainties in the normal-state background, or by a contribution from the larger in-plane signal coming from misalignments in the applied fields or currents.

The first aim of this article is to present quantitative experimental results on the FD in 2H-NbSe$_2$ single crystals for $H$ applied in the two main crystallographic directions. The moderate uniaxial anisotropy  of 2H-NbSe$_2$ ($\gamma\approx3$),\cite{Toyota} together with its accessible critical temperature ($T_C\approx 7.15$ K) and the possibility of growing large and very homogeneous single crystals, makes this material and excellent candidate to study the FD anisotropy. In fact, the interest of 2H-NbSe$_2$ to study superconducting fluctuations was early recognized by Prober, Beasley and Schwall (PBS).\cite{Prober77} However, in their pioneering work they just measured the in-plane FD and, in addition, they used a very low field amplitude ($\sim 1$ mT). As a consequence, even for this FD component it was not possible to separate the intrinsic FD from the effect of extrinsic $T_C$ inhomogeneities or from possible \textit{``unknown effects of the charge density wave"} (CDW) which appears below $T_{\rm CDW}\approx33$ K in this material (Section VI.B of Ref.~\onlinecite{Prober77}). 

To overcome the difficulties commented above, in our experiments we first use $H$ amplitudes ranging from 0.05 T to 5.5 T (i.e., up to $5\times10^3$ times the amplitudes used in Ref.~\onlinecite{Prober77}), which avoid $T_C$-inhomogeneity effects,\cite{nota} and cover both the low- and finite-field FD regimes. Moreover, to obtain the signal-to noise ratio needed to resolve the small transverse FD, we prepared a sample consisting in a stack of six large high-quality 2H-NbSe$_2$ single crystals. These different experimental improvements allowed us to measure the FD in the two main crystallographic directions, and to disentangle the nonlocal electrodynamic effects from the anisotropy influence. Our results also provide unambiguous answers to the questions opened by PBS.\cite{Prober77} In particular, we show that the FD is not affected by the CDW state, in agreement with a phenomenological estimate that takes into account the coupling between CDW and superconductivity. 

The paper is organized as follows: In Sec.~II we present the experimental details (samples preparation, characterization and measurement) and results. The theoretical background and data analysis is presented in Sec.~III, and the conclusions in Sec.~IV.

\section{Experimental details and results}

\subsection{Crystals' growth, characterization, and measurements procedure}

The 2H-NbSe$_2$ single crystals used in this work were grown by the chemical vapor transport method. Their crystal structure was confirmed to be in the 2H-phase by x-ray diffraction. These samples have a platelet geometry (typically $\sim$1 mm thick and $\sim$4 mm diameter) with mirror-like surfaces perpendicular to the $c$-axis. The crystals' volumes were deduced from their masses and the 2H-NbSe$_2$ theoretical density resulting from the crystallographic parameters (6.22 g/cm$^3$). 

The magnetization, $M$, was measured with a commercial (Quantum Design) SQUID magnetometer. 
Most of the measurements were performed on a stack of six single crystals (glued with a minute amount of GE varnish), with a total superconducting volume of 73.7 mm$^3$. As we will see below, this is particularly important for $H\parallel ab$ due to the small FD signal. For the measurements in this direction, the sample orientation was controlled to better than 0.5$^{\circ}$, which ensured that the FD contribution from the perpendicular direction was negligible (see below).

\subsection{Determination of the superconducting parameters relevant for the FD analysis}

As a first magnetic characterization of the single crystals' stack, in Fig.~1 we present the temperature dependence of the zero-field-cooled (ZFC) and field-cooled (FC) magnetic susceptibilities, measured with a magnetic field of 0.6 mT applied perpendicular to the crystallographic \textit{ab} planes. These measurements were corrected for demagnetizing effects by using as demagnetizing factor $D=0.40$, which is compatible with the stack geometry and gives the expected value of $\chi_{\perp}^{ZFC}=-1$ at low temperatures. The zero-field transition temperature, $T_{C0}\approx7.15$~K, and the transition width, $\Delta T_{C0}\approx0.15$~K, were estimated as the maximum and, respectively, the full-width at half-maximum of the $d\chi_\perp^{FC}/dT$ vs $T$ curve. This $\Delta T_C$ value is in part due to the use in the measurements of a finite magnetic field.\cite{Tcbroadening} Once subtracted this last contribution, the transition broadening due to $T_C$ inhomogeneities is estimated to be $\sim$75 mK. This attests the excellent quality of all the single crystals used in the stack, and will allow to explore the fluctuation-induced magnetization for reduced temperatures as low as $\varepsilon\equiv\ln(T/T_{C0})\approx2\times10^{-2}$ (see below).

\begin{figure}[t]
\includegraphics[scale=0.5]{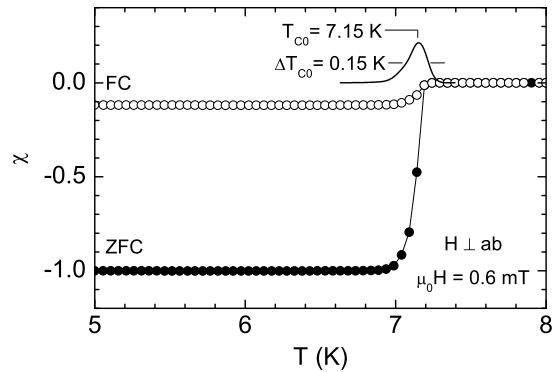}
\caption{Temperature dependence of the FC and ZFC magnetic susceptibilities of the single crystals' stack, obtained with a magnetic field of 0.6 mT applied perpendicular to the \textit{ab} planes. $T_{c0}$ and $\Delta T_{c0}$ were obtained from the maximum and, respectively, the width of the $d\chi_{\perp}^{FC}/dT$ vs. $T$ curve (represented as a solid line in arbitrary units). For details see the main text.} 
\label{FIG_Tc}
\end{figure}
\begin{figure}[t]
\includegraphics[scale=0.5]{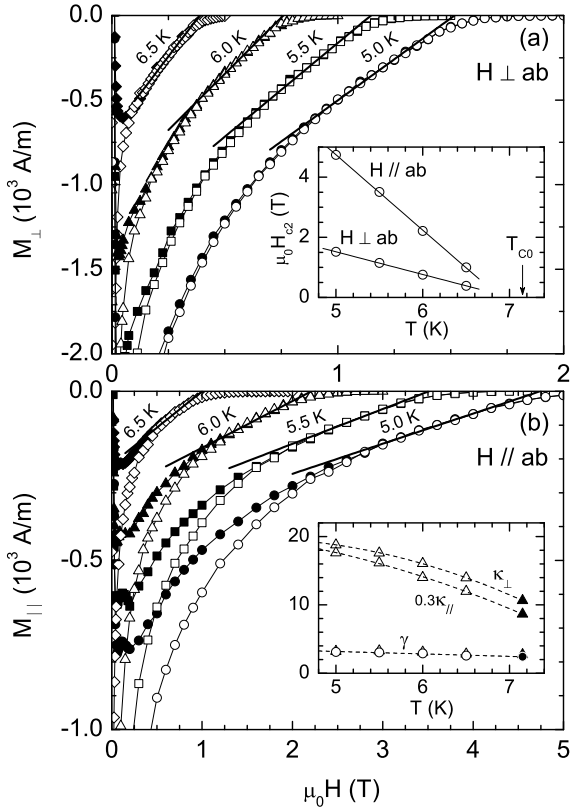}
\caption{Magnetization vs. external magnetic field curves obtained with $H$ applied in the main crystallographic directions at some constant temperatures below $T_C$. Open and closed symbols were obtained by increasing and decreasing $H$ respectively. The thin lines are guides for the eyes while the thick ones are fits of the Abrikosov's GL result to the data in the linear region near $H_{C2}(T)$. The resulting $H_{C2}^{\perp}$, $H_{C2}^{\parallel}$, $\kappa_\perp$, $\kappa_\parallel$ and $\gamma$ are shown in the insets.} 
\label{FIG_imanacion}
\end{figure}

Other superconducting parameters involved in the analysis of the fluctuation-induced diamagnetism (as the coherence length amplitude, the anisotropy factor or the GL parameter) were determined from the $H$ dependence of the the mixed state magnetization. These measurements were performed with both $H \perp ab$ and $H \parallel ab$ in a single crystal from the stack. The results are shown in Fig.~2. As it may be clearly seen, the magnetization is highly reversible in a broad region near the corresponding upper critical fields. Also, the so-called ``peak effect" (sometimes attributed to sample inhomogeneities\cite{peakeffect}) is completely absent from these curves. Both facts allow an easy comparison with the high-field Abrikosov's result, that for anisotropic superconductors may be approximated by\cite{Tinkham,Kogan81}
\begin{equation}
M_\perp(T,H)=\frac{H-H_{C2}^\perp(T)}{2\beta_A\kappa_\perp^2},
\end{equation}
and
\begin{equation}
M_\parallel(T,H)=\frac{H-H_{C2}^\parallel(T)}{2\beta_A\kappa_\parallel^2}.
\end{equation}
Here $H_{C2}^\perp$ and $H_{c2}^\parallel$ are the upper critical magnetic fields for $H\perp ab$ and $H\parallel ab$ respectively, $\kappa_\perp\equiv \lambda_{ab}/\xi_{ab}$ and $\kappa_\parallel=\gamma\kappa_\perp$ are the corresponding Ginzburg-Landau (GL) parameters ($\lambda_{ab}$ and $\xi_{ab}$ are, respectively, the in-plane magnetic penetration and coherence lengths) and $\beta_A$ is the Abrikosov parameter ($\sim 1.16$ for a triangular vortex lattice). In Eqs.~(1) and (2) we have ignored terms of the order of the unity in the denominator, which can be fully neglected with respect to $2\beta_A\kappa_{\perp,\parallel}^2$ when $\kappa_{\perp,\parallel}\gg1$. The thick lines in Fig.~2 are linear fits of Eqs.~(1) and (2) to the $M(H)_T$ curves in the reversible region. The resulting $H_{C2}^\perp$, $H_{C2}^\parallel$, $\kappa_\perp$ and $\kappa_\parallel$ are presented in the insets of that figure. As it may be clearly seen, in the temperature range studied (0.7 $\lesssim T/T_{C}\lesssim 0.9$) both upper critical fields are linear in $T$.\cite{nonlinearHC2} Linear extrapolation to $T=0$~K leads to $\mu_0H_{C2}^\perp(0)\approx5.3$~T and $\mu_0H_{C2}^\parallel(0)\approx17.3$~T. By combining these values with the well known GL expressions
\begin{eqnarray}
\mu_0H_{C2}^\perp(0)&=&\frac{\phi_0}{2\pi\xi_{ab}^2(0)}\\
\mu_0H_{C2}^\parallel(0)&=&\frac{\phi_0}{2\pi\xi_{ab}(0)\xi_c(0)}
\end{eqnarray}
we obtained an in-plane coherence length amplitude of $\xi_{ab}(0)=78.8$ \AA\; and a transverse coherence length amplitude of $\xi_c(0)=24.2$ \AA. This last value is much larger than the periodicity length of the layered structure ($\sim5$ \AA) and, then, a description of the fluctuation effects above $T_C$ in terms of three-dimensional anisotropic theoretical approaches will be justified. Both in-plane and transverse GL parameters are also temperature dependent and extrapolate to $\kappa_\perp\sim11$ and respectively $\kappa_\parallel\sim29$ at $T_C$ (solid triangles), a value close to the one of some high-temperature cuprate superconductors. The anisotropy factor is also presented in the inset of Fig. 2(b) as calculated from $\gamma=H_{C2}^\parallel/H_{C2}^\perp$ (circles) and from $\gamma=\kappa_\parallel/\kappa_\perp$ (triangles). The agreement between these two independent determinations of $\gamma$ is an important consistency check of the procedure used to obtain the 2H-NbSe$_2$ superconducting parameters. As it may be seen, $\gamma$ presents a mean value of $\sim3$ in the temperature range studied, and extrapolates to $\gamma\approx2.4$ at $T_{C0}$.

\begin{figure}[t]
\includegraphics[scale=0.5]{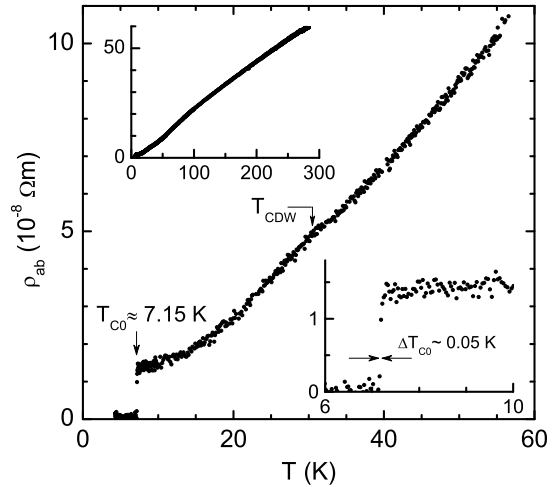}
\caption{Temperature dependence of the in-plane electrical resistivity. The kink at $\sim$30 K corresponds to the transition to the \textit{charge-density wave} state.} 
\label{FIG_ro}
\end{figure}

\begin{table}[b]
\caption{\label{tab:table1}2H-NbSe$_2$ superconducting parameters relevant for the FD analysis. They were obtained from the magnetization and resistivity measurement presented in Sec.~II.B.}
\begin{ruledtabular}
\begin{tabular}{cccccc}
 & $\mu_0H_{C2}^\perp(0)$ & $\kappa_\perp(T_{C0})$ & $\xi_{ab}(0)$ & & $\ell_{ab}$ \\
$T_{C0}$ & $\mu_0H_{C2}^\parallel(0)$ & $\kappa_\parallel(T_{C0})$ & $\xi_c(0)$ & $\gamma(T_{C0})$ & $\ell_c$\\
(K)   & (T) & & (\r{A}) &   & (\r{A}) \\
\hline
7.15 & 5.3 & 11 & 78.8 & 2.4 & 1830 \\
 & 17.3 & 29 & 24.2 & & $\sim$20\\
\end{tabular}
\end{ruledtabular}
\end{table}

In Fig.~\ref{FIG_ro} we present the temperature dependence of the in-plane electrical resistivity $\rho_{ab}$. These measurements were performed in a $2.2\times1.0\times0.067$ mm$^3$ sample cut from one of the single crystals used in the magnetization experiments, by using a \textit{van der Pauw} contacts configuration. The resulting $T_{C0}$ and $\Delta T_{C0}$ values are in excellent agreement with the ones obtained from the low-field magnetic susceptibility. The ratio $\rho_{ab}$(300 K)/$\rho_{ab}$(7.5 K) = 43, is close to the ones found in the best crystals.\cite{Lee69} The in-plane electronic mean free path close to $T_{C0}$, $\ell_{ab}\approx$ 1830 \r{A}, was estimated from the in-plane residual resistivity and from the carriers concentration\cite{Lee69} by using a Drude-model relation. The resistivity anisotropy is found to be $\rho_{c}/\rho_{ab}\sim 10^2$, which leads to an electronic mean free path in the $c$ direction of $\ell_{c}\sim$ 20 \r{A}. 

The 2H-NbSe$_2$ parameters obtained in this Section are summarized in Table I. These values are in excellent agreement with the ones that may be found in the literature.\cite{Toyota}

\subsection{Background extraction and fluctuation magnetization above $T_{C0}$}

An example of the as-measured $M(T)$ above $T_{C0}$ corresponding to a magnetic field of 0.5 T applied in both $H \perp ab$ and $H\parallel ab$ orientations is presented in the inset of Fig.~4(a). The difference of a factor $\sim$2 between both directions was already reported and explained in terms of the anisotropic electronic structure of the material.\cite{MacDonald81} The lines are the normal-state or background contributions, $M_B(T)_H$, obtained by fitting a Curie-like function [$M_B(T)_H=a+bT+c/T$] in the temperature region between $\sim11$ K ($\sim1.5T_{C}$) and $\sim20$ K ($\sim2.8T_{C}$). The lower limit corresponds to a temperature where the fluctuation effects are expected to be negligible,\cite{Vidal} while the upper limit is well below $T_{\rm CDW}$. As it may be already seen in this figure, the effect of the superconducting fluctuations is clearly anisotropic, being more pronounced for $H\perp ab$.

\begin{figure}[b]
\includegraphics[scale=0.5]{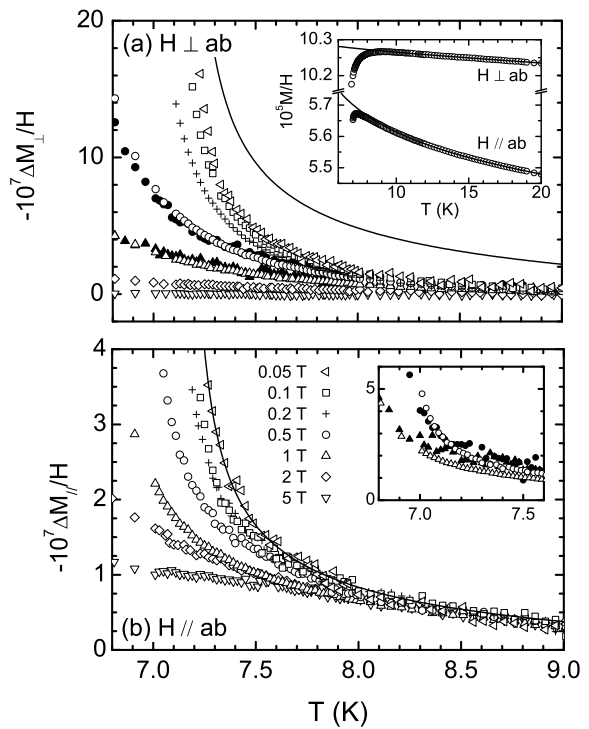}
\caption{Temperature dependence of the fluctuation magnetization (over $H$) for several magnetic fields applied in the two main crystallographic directions. For comparison, some data obtained in a reference single crystal are included (solid symbols). The lines correspond to the GGL approach in the low-$H$ limit [Eqs.(7) and (9)]. Inset in (a): Example of the normal-state background subtraction, corresponding to $\mu_0H=0.5$ T (see the main text for details). Inset in (b): Detail around $T_C$ including data obtained in the reference single crystal.} 
\label{FIG_DXvsT}
\end{figure}

In the main Fig.~4 we present the temperature dependence around $T_{C0}$ of the fluctuation-induced magnetization, $\Delta M\equiv M-M_B$, normalized by their corresponding $H$ amplitudes. These data were obtained in the single crystals' stack with $H\perp ab$ and $H\parallel ab$. For comparison, we also include some measurements performed in a 17.2 mm$^3$ reference single crystal from the stack (solid symbols). As shown in the inset of Fig.~4(b), when $H\parallel ab$ the only way to obtain an acceptable signal-to-noise ratio above $T_{C0}$ is by using the single crystals' stack. But also, wherever the experimental resolution is enough (typically for $\left|\Delta M/H\right|\stackrel{>}{_\sim}0.3\times10^{-6}$ when $\mu_0H=1-0.5$ T) the results in the single crystal match the ones in the stack. This is an important reliability check of the measurements in the stack which confirms the negligible effect of possible sample misalignments.

The results in Fig.~4 confirm that the saturation FD amplitude in the low-field limit [$H/H_{C2}^{\perp,\parallel}(0)\ll\varepsilon$, where $\varepsilon\equiv\ln(T/T_{C0})$ is the reduced temperature] is several times larger for $H\perp ab$ than for $H\parallel ab$. When the magnetic field is increased $-\Delta M_\perp/H$ is progressively reduced to end up vanishing when $H\sim H_{C2}^\perp(0)$. This effect has been interpreted in terms of the shrinkage, induced by the magnetic field, of the superconducting wave-function to lengths of the order of the Cooper pairs size perpendicular to the applied field.\cite{Vidal,Soto04} Due to the high $H_{C2}^\parallel(0)$ value ($\sim 17$ T) $\Delta M_\parallel/H$ does not vanish even for the highest magnetic fields used in the experiments. 

\section{Comparison with the Gaussian Ginzburg-Landau approach.}

\subsection{Theoretical background}

To analyze quantitatively the results of Fig.~4, as $\xi_c(0)$ is much larger than the layers periodicity length, we use the 3D anisotropic Gaussian Ginzburg-Landau (GGL) theory. Within this approach, the magnetization of a 3D anisotropic superconductor may be obtained by generalizing the result for 3D isotropic materials, $\Delta M_{\rm iso}$, through the scaling transformation: 
\begin{equation}
M(\varepsilon,h,\theta)=\gamma a(\theta)M_{\rm iso}(\varepsilon,ha(\theta)).
\label{scaling}
\end{equation}
Here $\theta$ is the angle between the external magnetic field and the $c$ crystallographic direction, $a(\theta)\equiv\sqrt{\cos^2\theta+\gamma^{-2}\sin^2\theta}$ and $h\equiv H/H_{C2}^\perp(0)$ the reduced magnetic field. This transformation was introduced by Klemm and Clem,\cite{Klemm80} and generalized by Blatter\cite{Blatter92} and by Hao and Clem\cite{Hao92} to different observables and different regions in the $H-T$ phase diagram. In particular, it was shown to be also valid for the fluctuation region above $T_{C0}$.\cite{Blatter92} Buzdin and Feinberg\cite{Buzdin94} used Eq.~(5) to obtain a $\Delta M(\varepsilon,h,\theta)$ expression, but their result does not take into account the limits imposed by the uncertainty principle to the shrinkage of the superconducting wavefunction at high $\varepsilon$ or $h$.\cite{Vidal} In Ref.~\onlinecite{Mosqueira01} it was obtained an expression for $\Delta M_{\rm iso}(\varepsilon,h)$ valid for finite reduced magnetic fields and under a \textit{total-energy} cutoff to include this quantum constraint. By combining that result with Eq.~(\ref{scaling}), the fluctuation magnetization for $H\perp ab$ ($\theta=0$) is given by
\begin{widetext}
\begin{eqnarray}
\Delta M_\perp(\varepsilon,h)&=&-\frac{k_BT\gamma\sqrt{2h}}{\pi\phi_0\xi_{ab}(0)}\int_{0}^{\sqrt{(c-\varepsilon)/2h}}dx\left[\frac{c-\varepsilon}{2h}-\left(\frac{c}{2h}+x^2\right)\psi\left(\frac{c+h}{2h}+x^2\right)+\right.\nonumber\\
&+&\left.\ln\Gamma\left(\frac{c+h}{2h}+x^2\right)+\left(\frac{\varepsilon}{2h}+x^2\right)\psi\left(\frac{\varepsilon+h}{2h}+x^2\right)-\ln\Gamma\left(\frac{\varepsilon+h}{2h}+x^2\right)\right],
\label{Mperp}
\end{eqnarray}
where $\Gamma$ and $\psi$ are, respectively, the gamma and digamma functions, and $c\approx0.5$ is the cutoff constant.\cite{Vidal,Mosqueira01} In the low magnetic field limit ($h\ll\varepsilon$), this equation is linear in $h$ and reduces to
\begin{equation}
\frac{\Delta M_\perp}{H}(\varepsilon)=-\frac{k_BT\mu_0\gamma\xi_{ab}(0)}{3 \phi_{0}^2}
\left[\frac{\arctan\sqrt{(c-\varepsilon)/\varepsilon}}{\sqrt{\varepsilon}}
-\frac{\arctan\sqrt{(c-\varepsilon)/c}}{\sqrt{c}}
\right].
\label{Xperp}
\end{equation}
\end{widetext}

In absence of any cutoff (i.e., when $c\to\infty$) it further simplifies to $\Delta M_\perp/H=-\pi k_BT\mu_0\gamma\xi_{ab}(0)/6\phi_{0}^2\sqrt{\varepsilon}$, which is the well known Schmidt and Schmid result.\cite{Schmidt,Schmid} The fluctuation magnetization for $H\parallel ab$ ($\theta=\pi/2$) for finite magnetic fields is found to be
\begin{equation}
\Delta M_\parallel(\varepsilon,h)=\frac{1}{\gamma}\Delta M_\perp\left(\varepsilon,\frac{h}{\gamma}\right),
\label{Mpara}
\end{equation}
where $\Delta M_\perp$ is given by Eq.(\ref{Mperp}). In turn, in the low magnetic field limit ($h\ll\varepsilon$) it reduces to
\begin{equation}
\frac{\Delta M_\parallel}{H}(\varepsilon)=\frac{1}{\gamma^2}\frac{\Delta M_{\perp}}{H}(\varepsilon),
\label{Xpara}
\end{equation}
where $\Delta M_\perp/H$ is now given by Eq.(\ref{Xperp}).

\subsection{Data analysis}

The lines in Figs.~4(a) and 4(b) are the GGL result for $\Delta M$ in the low magnetic field limit. They were obtained by using Eqs.(\ref{Xperp}) and (\ref{Xpara}) with the $T_{C0}$, $\xi_{ab}(0)$ and $\gamma$ values in Table I. For $H\parallel ab$ the theory is in excellent agreement with the data obtained under the lowest magnetic field. This agreement extends also to measurements performed under higher magnetic fields, although in a narrower temperature range (typically for $\varepsilon\stackrel{>}{_\sim}2h$, in accordance with the low-magnetic field condition). On the contrary, for $H\perp ab$ the GGL prediction overestimates by a factor $\sim$2 even the measurements under the lowest magnetic field. 

\begin{figure}[t]
\includegraphics[scale=0.5]{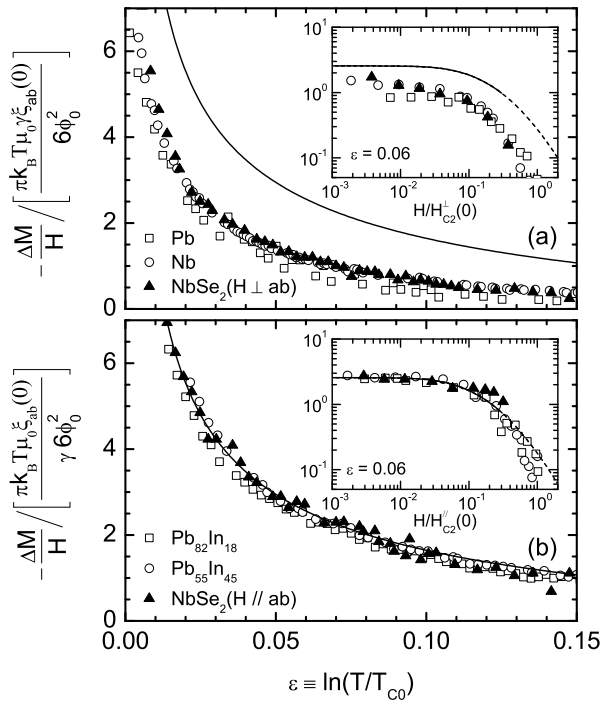}
\caption{Comparison of the FD in NbSe$_2$ with those of clean Pb and Nb (a) and dirty Pb-In alloys (b). Note the normalization by the corresponding \textit{Schmidt amplitude}. In the main figures it is presented the $\varepsilon$-dependence in the low-field limit ($H\sim10^{-2} H_{C2}^{\perp}$ and, respectively, $H\sim10^{-2} H_{C2}^{\parallel}$), and in the insets the $h$-dependence for $\varepsilon=0.06$. The lines are the GGL predictions (applicable only up to $H/H_{C2}^{\perp,\parallel}\stackrel{<}{_\sim}0.3$) [Eqs.(6) and (8)]. Nevertheless, these data show that the FD in both directions vanishes at $H\approx H_{C2}^{\perp,\parallel}$, in agreement with the total energy cutoff prediction.\cite{Vidal,Soto04}} 
\label{FIG_comparacion}
\end{figure}

The adequacy of the GGL theory to explain the FD of 2H-NbSe$_2$ when $H\parallel ab$, is typical of dirty metallic alloys like Pb$_{1-x}$In$_x$ with $0.08\geq x\geq0.45$,\cite{Soto04,Mosqueira01,SotoPr,Vidal} and of HTSC\cite{DX_HTSC} which may also be considered as dirty materials.\cite{Homes05} This is consistent with the fact that for electric currents flowing in the $c$ direction (which are induced when $H\parallel ab$) the electronic mean free path $\ell_c$ is close to $\xi_c(0)$ (see Table I). On the other side, to test whether the features observed in the FD of 2H-NbSe$_2$ when $H\perp ab$ are due to nonlocal electrodynamic effects, as in clean or moderately dirty low-$T_C$ superconductors,\cite{Gollub,Kurkijarvi72,Mosqueira03} in Fig.~5(a) we compare these measurements with the ones in pure Pb and Nb. The differences in the $T_C$, $\xi(0)$ and $\gamma$ values of these compounds were compensated by scaling $\Delta M/H$ by the corresponding \textit{Schmidt amplitude} [$\pi k_BT\mu_0\gamma a^2(\theta)\xi_{ab}(0)/6\phi_{0}^2$], and by using reduced temperatures and magnetic fields. As may be clearly seen, the coincidence between the scaled $\Delta M/H$ in these three compounds, in both the $\varepsilon$- and $h$-dependences, suggests that 2H-NbSe$_2$ could be affected (for $H\perp ab$) by nonlocal effects in the same way as pure elements. This is consistent with the fact that for electric currents flowing in the $ab$ planes (the ones induced when $H\perp ab$) $\ell_{ab}\gg\xi_{ab}(0)$ (see Table I), and 2H-NbSe$_2$ is well in the clean limit. For completeness, in Fig.~5(b) we compare the results in 2H-NbSe$_2$ when $H\parallel ab$ with the ones in some dirty Pb-In alloys. As may be clearly seen in this figure, the GGL approach is in excellent agreement with the measurements in the accessible temperature and magnetic fields ranges.

The results summarized above seem to discard a contribution to the FD coming from the coupling between superconductivity and CDW in this compound. An estimate of this contribution may be obtained on the grounds of the semi-phenomenological GL-functional proposed in Ref.~\onlinecite{Machida81}. For $T_{\rm CDW}\gg T_C$ (a case well adapted for 2H-NbSe$_2$) this functional just differs from the conventional one by a coupling term $K\cos{(\vec{Q}\cdot\vec{r})}$, where $K$ is a dimensionless constant which amplitude is expected to decrease with $T_C/T_{\rm CDW}$, and $\vec{Q}$ is the CDW wavevector. For 2H-NbSe$_2$ $Q\sim 2k_{F}\sim 10^{10}$ m$^{-1}$,\cite{Machida81} and since $K\sim 0.1$,\cite{balseiro} the coupling can be treated perturbatively. The main correction is expected to happen for $H\perp ab$ and it may be approached by renormalizing the reduced temperature as
\begin{equation}
\varepsilon\to\varepsilon+K \exp\left[-\frac{Q^{2}\xi^2(0)}{4h}\right].
\end{equation}
As $Q\gg\xi^{-1}(0)$ for 2H-NbSe$_2$, Eq.~(10) predicts negligible effects of the CDW on the FD in this superconductor for the experimentally accessible $\varepsilon$ and $h$ values, in agreement with our measurements. However, these effects may play a non negligible role close to $T_C$ in compounds with a smaller $\xi(0)$ and longer-wavelength CDW. This could be the case of the intrinsically inhomogeneous underdoped HTSC,\cite{McElroy} for which Eq.~(10) is valid whenever $T_C/T^*\ll1$, where $T^*$ is the pseudogap temperature which may be identified to $T_{\rm CDW}$.\cite{Klemm00}

\section{Conclusions}

The experimental results and analysis presented here answer the long-standing problem of the interrelation between anisotropy and nonlocal electrodynamic effects on the FD, an aspect which directly concerns the precursor superconductivity in the layered HTSC. In particular, for $H\parallel ab$ the FD above $T_C$ in 2H-NbSe$_2$ may be fully understood in terms of the GGL approach under a total-energy cutoff. For $H\perp ab$ the FD is found to be affected by nonlocal electrodynamic effects in the same extent as the clean low-$T_C$ isotropic superconductors, in agreement with the fact that $\ell_{ab}\gg\xi_{ab}(0)$ (Table I). Our results also discard a possible contribution to the FD coming from the CDW state in this compound. A phenomenological estimate of the interaction between CDW and superconductivity justifies this finding. In addition, our preset results suggest that the possible presence of a CDW state in underdoped HTSC\cite{Klemm00} could affect their FD due to the intrinsically inhomogeneous nature of these superconductors (which could lead to charge-density variations with characteristic lengths up to $\sim$3 nm)\cite{McElroy}. The observation of these effects will provide a direct experimental demonstration of the relevance of the CDW in these compounds.  

\section{Acknowledgments}

This work was supported by the Spanish Ministerio de Educaci\'on y Ciencia and FEDER funds (MAT2004-04364), and by the Xunta de Galicia (PGIDIT04TMT206002PR). H.B. and D.P. acknowledge the ongoing support by the EPFL and the Swiss National Science Foundation, F.S. by Union Fenosa, and C.C. by the Xunta de Galicia through the Isidro Parga-Pondal Programme.


\begin{references}
\bibitem{nota}As near $T_{C0}$, $|dT_C/d(\mu_0H)|\sim$ 1 K/T, these field amplitudes introduce a $T_C$ shift large enough to avoid the effect of $T_C$-inhomogeneities above $T_{C0}$.\cite{Cabo06} This simple procedure could not be used by PBS due to the limitations of their magnetometer.\cite{Prober77}
\bibitem{Tcbroadening}The 0.6 mT field used in the measurements shifts the temperature at which the sample enters in the Meissner state, $T_{C1}^{\perp}(H)$, by $\sim H/(dH_{C1}^\perp/dT)_{T_{C0}}\approx75$ mK, where $H_{C1}^\perp=H_{C2}^\perp\ln\kappa_\perp/2\kappa_\perp^2$ is the lower critical magnetic field for $H\perp ab$. This shift leads to a transition broadening of the same extent because between $T_{C1}^\perp(H)$ and $T_{C2}^\perp(H)\approx T_{C0}$ the samples passes through the mixed state. This shift is more pronounced when $H\parallel ab$, due to the smaller lower critical field for this direction.
\end{references}
\end{document}